\documentclass[conference,10pt]{IEEEtran}
\IEEEoverridecommandlockouts
\usepackage{cite}
\usepackage{amsmath,amssymb,amsfonts}
\usepackage{algorithmic}
\usepackage{graphicx}
\usepackage{textcomp}
\usepackage{etoolbox}
\usepackage{float}

\floatstyle{ruled}
\newfloat{algorithm}{tbp}{loa}
\providecommand{\algorithmname}{Algorithm}
\floatname{algorithm}{\protect\algorithmname}

\def\BibTeX{{\rm B\kern-.05em{\sc i\kern-.025em b}\kern-.08em
    T\kern-.1667em\lower.7ex\hbox{E}\kern-.125emX}}

    \makeatletter
\patchcmd{\@maketitle}
  {\addvspace{0.5\baselineskip}\egroup}
  {\addvspace{-1\baselineskip}\egroup}
  {}
  {}

\makeatother

\begin{document}

\title{Inter-Tenant Cooperative Reception for C-RAN Systems With Spectrum Pooling}

\author{\IEEEauthorblockN{Junbeom Kim, Daesung Yu, Seok-Hwan Park} \IEEEauthorblockA{\textit{\small Dept. of Elect. Engineering} \\
 \textit{Jeonbuk National University}\\
 Jeonju, Korea \\
 \{junbeom, imcreative93, seokhwan\}@jbnu.ac.kr} \and \IEEEauthorblockN{Osvaldo Simeone} \IEEEauthorblockA{\textit{\small KCLIP Lab, Centre for Telecomm. Research} \\
 \textit{\small Dept. of Engineering} \\
 \textit{King's College London}\\
 London, UK \\
 osvaldo.simeone@kcl.ac.uk} \and \IEEEauthorblockN{Shlomo Shamai (Shitz)} \IEEEauthorblockA{\textit{\small Dept. of Elect. Engineering} \\
 \textit{Technion}\\
 Haifa, Israel \\
sshlomo@ee.technion.ac.il} }
\maketitle
\begin{abstract}
This work studies the uplink of a multi-tenant cloud radio access network (C-RAN) system with spectrum pooling. In the system, each operator has a cloud processor (CP) connected to a set of proprietary radio units (RUs) through finite-capacity fronthaul links.
The uplink spectrum is divided into private and shared subbands, and all the user equipments (UEs) of the participating operators can simultaneously transmit signals on the shared subband. To mitigate inter-operator interference on the shared subband, the CPs of the participating operators can exchange compressed uplink baseband signals on finite-capacity backhaul links. This work tackles the problem of jointly optimizing bandwidth allocation, transmit power control and fronthaul compression strategies. In the optimization, we impose that the inter-operator privacy loss be limited by a given threshold value. An iterative algorithm is proposed to find a suboptimal solution based on the matrix fractional programming approach. Numerical results validate the advantages of the proposed optimized spectrum pooling scheme.\end{abstract}


\section{Introduction\label{sec:intro}}

\let\thefootnote\relax\footnotetext{This work was supported by Basic Science Research Program through the National Research Foundation of Korea grants funded by the Ministry of Education [NRF-2018R1D1A1B07040322, NRF-2019R1A6A1A09031717]. The work of O. Simeone was supported by the European Research Council (ERC) under the European Union's Horizon 2020 research and innovation programme (grant agreement No 725731). The work of S. Shamai was supported by the ERC under the European Union's Horizon 2020 research and innovation programme (grant agreement No 694630).}

%by U.S. NSF through grant 1525629 and

%The work of S. Shamai has been supported by the European Union’s Horizon 2020 research and innovation programme, grant agreement no. 694630.

Network slicing is a key technology for future wireless communication systems \cite{Foukas-et-al:CM17}. Two examples of network slicing are radio access network (RAN) sharing and spectrum pooling, in which network operators share infrastructure nodes or frequency spectrum in order to meet the growing demands for high data rates \cite{Boccardi-et-al:JCN16,Aydin-et-al:TWC17}. Another promising network architecture is cloud RAN (C-RAN), which is being deployed for performance evaluation. In a C-RAN system, baseband signal processing on behalf of a set of distributed radio units (RUs), also known as distributed units (DUs) in 5G New Radio (NR) \cite{ITU-T:18},
is jointly carried out by a cloud processor (CP), known as central unit (CU) in 5G NR \cite{ITU-T:18}, that is connected to the RUs through fronthaul links \cite{Park-et-al:TVT13,Park-et-al:TVT16,Hong-et-al:TWC16}.
Reference \cite{Park-et-al:TVT18} studied the downlink of a multi-tenant C-RAN system with spectrum pooling, in which C-RAN downlink systems of two network operators cooperate to maximize the total throughput.

\begin{figure}
\centering
\!\!\!\!\!\!\!\!\centerline{\includegraphics[width=8.2cm, height=7cm]{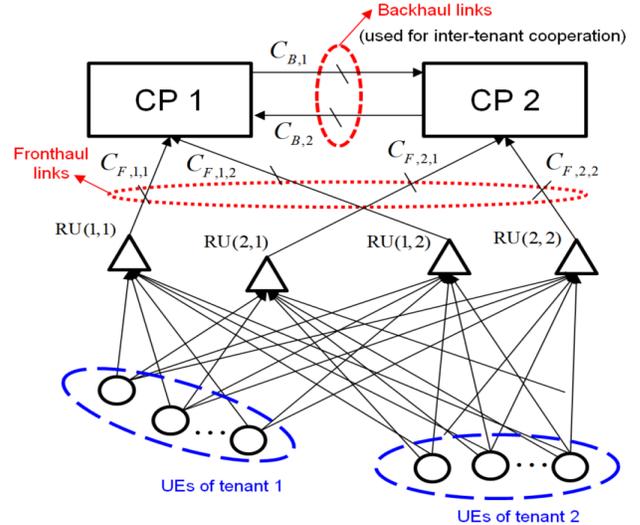}}
\caption{Illustration of the uplink of a multi-tenant C-RAN system.}
\label{fig:system-model}
\end{figure}

In this work, we focus instead on the uplink of a multi-tenant C-RAN system with spectrum pooling. As in \cite{Park-et-al:TVT18}, we assume that the uplink spectrum is divided into private and shared subbands, and that all the user equipments (UEs) of the participating operators can simultaneously transmit uplink signals on the shared subband.
The CPs of different operators can exchange compressed uplink baseband signals on finite-capacity backhaul links in order to mitigate the impact of inter-operator interference signals on the shared subband.
We address the problem of jointly optimizing bandwidth allocation, transmit power control and fronthaul compression strategies with the goal of maximizing the sum-rate of all the UEs in the participating operators.
In the optimization, we impose constraints on transmit power, fronthaul and backhaul capacity, as well as on the amount of information leaked on backhaul links, since exchange of information on backhaul links may cause privacy loss among operators. To find an effective, but suboptimal, solution, we propose an iterative algorithm based on the matrix fractional programming (FP) approach \cite{Shen-et-al:TN19}. We observe that the tools used here are hence different from \cite{Park-et-al:TVT18}, which relied on successive convex approximation.

\section{System Model}\label{sec:system}

As shown in Fig. \ref{fig:system-model}, we consider the uplink of a multi-tenant C-RAN system with $N_O=2$ network operators. In the network of each operator $i$, $N_{U,i}$ single-antenna UEs send messages to the serving CP $i$ through $N_{R,i}$ RUs, where the $r$th RU of operator $i$ is connected to CP $i$ through a fronthaul link of capacity $C_{F,i,r}$ bits per second (bps).
We refer to the $k$th UE and the $r$th RU of the $i$th operator as UE $(i,k)$ and RU $(i,r)$, respectively, and denote the number of antennas of RU $(i,r)$ by $n_{R,i,r}$. We also define the sets
$\mathcal{N}_{O}=\{1,2\}$, $\mathcal{N}_{R,i}=\{1, \ldots, N_{R,i}\}$ and $\mathcal{N}_{U,i}=\{1, \ldots, N_{U,i}\}$.
As in \cite{Park-et-al:TVT18}, we assume that CP $i$ can send information to the other CP $\bar{i}$ on a backhaul link of capacity $C_{B,i}$ bps, where $\bar{i}$ denotes $\bar{i}=3-i$ for $i\in \mathcal{N}_O$.

To enable spectrum sharing among tenants, we partition the uplink spectrum of bandwidth $W$ [Hz] into three subbands: two private subbands $i\in N_O$ of bandwidth $W_{P,i}$ and a shared subband of bandwidth $W_S$. The bandwidth parameters $\mathbf{W}=\{W_{P,i}\}_{i\in\mathcal{N}_O}\cup \{W_S\}$ must satisfy the constraint $\sum_{i\in \mathcal{N}_{O}}W_{P,i}+W_{S} = W$.
Traditional C-RAN uplink systems without spectrum pooling \cite{Park-et-al:TVT13} can be modeled by fixing private bandwidths $W_{P,i}$ and by setting $W_S=0$. In this work, we include the bandwidth allocation variables $\mathbf{W}$ into the design space to maximize the spectral efficiency.

We model the signal $\mathbf{y}_{i,r}^{(i)}\in\mathbb{C}^{n_{R,i,r}\times 1}$ received by RU $(i,r)$ on its private subband $i$ as
\begin{equation}
\mathbf{y}_{i,r}^{(i)}=\sum\nolimits_{k\in\mathcal{N}_{U,i}}\mathbf{h}_{i,k}^{i,r}x_{i,k}^{(i)}+\mathbf{z}_{i,r}^{(i)},\label{eq:received-signal-private}
\end{equation}
where $\mathbf{h}_{j,k}^{i,r}\in\mathbb{C}^{n_{R,i,r}\times 1}$ denotes the channel vector from UE $(j,k)$ to RU $(i,r)$; $x_{i,k}^{(i)}$ represents the transmit signal of UE $(i,k)$ on the private subband; and $\mathbf{z}_{i,r}^{(i)}\sim\mathcal{CN}(\mathbf{0},N_{0}\mathbf{I})$ is the additive noise vector at RU $(i,r)$ on the private subband.
We impose transmit power constraints $\mathtt{E}[|x_{i,k}^{(i)}|^2]\leq P_{\max}$ for all $i\in\mathcal{N}_O$ and $k\in\mathcal{N}_{U,i}$.

In a similar way, the signal $\mathbf{y}_{i,r}^{(S)}\in\mathbb{C}^{n_{R,i,r}\times 1}$ received by RU $(i,r)$ on the shared subband is modeled as
\begin{equation}
\mathbf{y}_{i,r}^{(S)}=\sum\nolimits_{k\in\mathcal{N}_{U,i}}\! \mathbf{h}_{i,k}^{i,r}x_{i,k}^{(S)}+\sum\nolimits_{k\in\mathcal{N}_{U,\bar{i}}}\!\mathbf{h}_{\bar{i},k}^{i,r}x_{\bar{i},k}^{(S)} + \mathbf{z}_{i,r}^{(S)},
\label{eq:received-signal-shared}
\end{equation}
where $x_{i,k}^{(S)}$ denotes the signal transmitted by UE $(i,k)$ on the shared subband which satisfies the transmit power constraint $\mathtt{E}[|x_{i,k}^{(S)}|^2]\leq P_{\max}$, and $\mathbf{z}_{i,r}^{(S)}\sim\mathcal{CN}(\mathbf{0},N_{0}\mathbf{I})$ is the additive noise at RU $(i,r)$ on the shared subband. We note that the second term in the right-hand side (RHS) of (\ref{eq:received-signal-shared}) represents the interference signal caused by the uplink transmission of the other tenant's UEs on the shared subband.

\section{Proposed Multi-Tenant C-RAN Uplink System}\label{sec:multi-tenant}

In this section, we describe the operations of UEs, RUs, and CPs through inter-tenant cooperative uplink reception.

\subsection{Channel Encoding and Power Control\label{subsec:encoding}}

Each UE $(i,k)$ splits its message $M_{i,k}$ of rate $R_{i,k}$ into two submessages $M_{i,k}^{(i)}$ and $M_{i,k}^{(S)}$ of rates $R_{i,k}^{(i)}$ and $R_{i,k}^{(S)}$, respectively, with $R_{i,k}^{(i)}+R_{i,k}^{(S)} = R_{i,k}$. The messages $M_{i,k}^{(i)}$ and $M_{i,k}^{(S)}$ are communicated on the private and shared subbands, respectively.

UE $(i,k)$ encodes each split submessage $M_{i,k}^{(m)}$, $m\in\{i,S\}$, with a Gaussian channel codebook. The encoded baseband signal denoted as $s_{i,k}^{(m)}$ is then distributed as $s_{i,k}^{(m)}\sim\mathcal{CN}(0,1)$.
The transmitted signal $x_{i,k}^{(m)}$ of UE $(i,k)$ on the private ($m=i$) or shared subband ($m=S$) is given as
\begin{equation}
x_{i,k}^{(m)} = v_{i,k}^{(m)} s_{i,k}^{(m)},
\end{equation}
where $v_{i,k}^{(m)}$ controls of the power of the signal $x_{i,k}^{(m)}$ and is subject to the constraint $0\leq |v_{i,k}^{(m)}|^2 \leq P_{\max}$.

\subsection{Fronthaul Quantization\label{subsec:quantization}}

RU $(i,r)$ forwards information on the uplink received signals $\mathbf{y}_{i,r}^{(m)}$, $m\in\{i,S\}$, to the CP for decoding. Since the fronthaul link that carries the signals $\mathbf{y}_{i,r}^{(m)}$ has a finite capacity $C_{F,i,r}$ bps, RU $(i,r)$ sends a quantized information of the signals. Following a standard approach \cite{Park-et-al:TVT13,Park-et-al:TVT16,Park-et-al:TVT18} for fronthaul quantization, we adopt the Gaussian test channel so that a quantized description of $\mathbf{y}_{i,r}^{(m)}$, denoted as $
\hat{\mathbf{y}}_{i,r}^{(m)}$, is modeled as
\begin{equation}
\label{eq:compression}
\mathbf{\hat{y}}_{i,r}^{(m)}=\mathbf{L}_{i,r}^{(m)}\mathbf{y}_{i,r}^{(m)}+\mathbf{q}_{i,r}^{(m)},
\end{equation}
where $\mathbf{L}_{i,r}^{(m)}\in\mathbb{C}^{n_{R,i,r}\times{n_{R,i,r}}}$ is a linear transformation matrix applied before the quantization, and $\mathbf{q}_{i,r}^{(m)}$ represents the quantization noise signal which is independent of the signal $\mathbf{y}_{i,r}^{(m)}$ and distributed as $\mathbf{q}_{i,r}^{(m)}\sim\mathcal{CN}(\mathbf{0},\mathbf{\Omega}_{i,r}^{(m)})$. We set the quantization noise covariance matrix to $\mathbf{\Omega}_{i,r}^{(m)}
=\mathbf{I}$ without loss of optimality, which suggests that we control the fronthaul quantization strategy by the choice of the transformation matrix $\mathbf{L}_{i,r}^{(m)}$.

The compression rates needed to express the quantized signals $\mathbf{\hat{y}}_{i,r}^{(i)}$ and $\mathbf{\hat{y}}_{i,r}^{(S)}$ are given as $W_{P,i}I(\mathbf{y}_{i,r}^{(i)}; \mathbf{\hat{y}}_{i,r}^{(i)})$ and $W_{S}I(\mathbf{y}_{i,r}^{(S)}; \mathbf{\hat{y}}_{i,r}^{(S)})$, respectively, where the mutual information values are computed as
\begin{align}
& I\left(\mathbf{y}_{i,r}^{(i)};\mathbf{\hat{y}}_{i,r}^{(i)}\right)\triangleq g_{i,r}^{(i)}\left(\mathbf{L}_{i,r}^{(i)},\mathbf{v}_{i}^{(i)}\right)\label{eq:quantization}\\
& =  \mathrm{log}_{2}\mathrm{det}\left(\sum\nolimits_{l\in\mathcal{N}_{U,i}} \! \lambda\left(\mathbf{L}_{i,r}^{(i)}\mathbf{h}_{i,l}^{i,r}v_{i,l}^{(i)}\right)+N_{0}\lambda\left(\mathbf{L}_{i,r}^{(i)}\right)+\mathbf{I}\right), \nonumber \\
& I\left(\mathbf{y}_{i,r}^{(S)};\mathbf{\hat{y}}_{i,r}^{(S)}\right)\triangleq g_{i,r}^{(S)}\left(\mathbf{L}_{i,r}^{(S)},\mathbf{v}_{i}^{(S)},\mathbf{v}_{\bar{i}}^{(S)}\right)\label{eq:quantization}\\
& =  \mathrm{log}_{2}\mathrm{det}\left(\begin{array}{c}\!\sum_{j\in\mathcal{N}_O}\sum_{l\in\mathcal{N}_{U,j}}\lambda\left(\mathbf{L}_{i,r}^{(S)}\mathbf{h}_{j,l}^{i,r}v_{j,l}^{(S)}\right)\!\\ +N_{0}\lambda\left(\mathbf{L}_{i,r}^{(S)}\right)+\mathbf{I}\!\end{array}\right), \nonumber
\end{align}
where we have defined the notations
$\mathbf{v}_{i}^{(m)}=\{v_{i,k}^{(m)}\}_{k\in\mathcal{N}_{U,i}}$,  $m\in\{i,S\}$, and $\lambda(\mathbf{A})=\mathbf{A}\mathbf{A}^{H}$.

The capacity constraint for the fonthaul link from RU $(i,r)$ to CP $i$ can be written as
\begin{equation}
\label{eq:fronthaul-constraint}
W_{P,i}g_{i,r}^{(i)}\left(\mathbf{L}_{i,r}^{(i)},\mathbf{v}_{i}^{(i)}\right)+W_{S}g_{i,r}^{(S)}\left(\mathbf{L}_{i,r}^{(S)},\mathbf{v}_{i}^{(S)},\mathbf{v}_{\bar{i}}^{(S)}\right) \leq C_{F,i,r},
\end{equation}
for $i\in\mathcal{N}_{O}$ and $r\in\mathcal{N}_{R,i}$.

\subsection{Multiplex-and-Forward Relaying for Inter-CP Cooperation}

The quantized signals $\hat{\mathbf{y}}_{i,r}^{(S)}$, which have been received on the shared subband and forwarded to CP $i$, contain interference from the UEs of the other operator $\bar{i}$. Therefore, the decoding of the signals $s_{i,k}^{(S)}$ transmitted on the shared subband can be potentially improved by exchanging the quantized signals $\hat{\mathbf{y}}_{i,r}^{(S)}$ among the CPs on the backhaul links.
We assume that CP $i$ forwards (a subset of) the bit streams received on the fronthaul link to the other CP $\bar{i}$ without any processing. This approach can be interpreted as an instance of multiplex-and-forward (MF) relaying strategy that was studied in \cite{Park-et-al:TVT16} for a multi-hop fronthaul network.

If the capacity $C_{B,i}$ of the backhaul link from CP $i$ to the other CP $\bar{i}$ is sufficiently large, it is desirable for CP $i$ to forward all
the quantized signals $\{\hat{\mathbf{y}}_{i,r}^{(S)}\}_{r\in\mathcal{N}_{R,i}}$ to the other CP $\bar{i}$ to maximize the impact of inter-CP cooperation.
However, if the capacity $C_{B,i}$  is relatively small, forwarding many signals $\{\hat{\mathbf{y}}_{i,r}^{(S)}\}_{r\in\mathcal{N}_{R,i}}$ on the backhaul link may limit the compression rates (i.e., the resolution of the quantized signals), leading to performance degradation.

Based on the above observation, we assume that CP $i$ sends a subset $\{\mathbf{{\hat{y}}}_{i,r}^{(S)}\}_{r\in\mathcal{S}_{R,i}}$ of the quantized signals to CP $\bar{i}$, where $\mathcal{S}_{R,i}$ is a subset of $\mathcal{N}_{R,i}$ with cardinality $|\mathcal{S}_{R,i}| = S_{R,i}$: If we set $S_{R,i} = N_{R,i}$, CP $i$ sends all the received quantized signals to CP $\bar{i}$, and if $S_{R,i} = 0$, CP $i$  sends no information on the backhaul link.
For the case of $1\leq S_{R,i} < N_{R,i}$, without claim of optimality, we assume that CP $i$ chooses the $S_{R,i}$ RUs corresponding to the largest channel magnitudes $|| \mathbf{h}_{\bar{i}}^{i,r} ||$ from the UEs of tenant $\bar{i}$, where $\mathbf{h}_{\bar{i}}^{i,r}\in\mathbb{C}^{n_{R,i,r}N_{U,\bar{i}}\times 1}$ is obtained by stacking the channel vectors from all the UEs $(\bar{i},k)$, $k\in\mathcal{N}_{U,\bar{i}}$, to RU $(i,r)$, i.e., $\mathbf{h}_{\bar{i}}^{i,r}\triangleq [\mathbf{h}_{\bar{i},1}^{i,r H} \mathbf{h}_{\bar{i},2}^{i,r H} \cdots \mathbf{h}_{\bar{i},N_{U,\bar{i}}}^{i,r H}]^H$.
Throughout the paper, we assume that the numbers $\{S_{R,i}\}_{i\in\mathcal{N}_O}$ are predetermined, and leave their optimization to future work.

For given subset $\mathcal{S}_{R,i}$,
the backhaul capacity constraint from CP $i$ to CP $\bar{i}$ can be stated as
\begin{equation}
\label{eq:fronthaul-constraint}
W_{S}\sum\nolimits_{r\in\mathcal{S}_{R,i}} \! g_{i,r}^{(S)}(\mathbf{L}_{i,r}^{(S)},\mathbf{v}_{i}^{(S)},\mathbf{v}_{\bar{i}}^{(S)})\leq C_{B,i}.
\end{equation}

\subsection{Decoding and  Achievable Rates\label{subsec:decoding}}

CP $i$ decodes the private-subband messages $M_{i,k}^{(i)}$, $k\in\mathcal{N}_{U,i}$, using the quantized signals $\mathbf{\hat{y}}_{i}^{(i)}=[\mathbf{\hat{y}}_{i,1}^{(i)H}\cdots\mathbf{\hat{y}}_{i,N_{R,i}}^{(i)H}]^{H}$ of the private subband that have been collected only from the proprietary RUs $(i,r)$, $r\in\mathcal{N}_{R,i}$. Assuming single-user detection (SUD), the achievable rate $R_{i,k}^{(i)}$ of message $M_{i,k}^{(i)}$ is given as $R_{i,k}^{(i)}=W_{P,i}I(s_{i,k}^{(i)};\mathbf{\hat{y}}_{i}^{(i)})$, where
 the mutual information can be written as
 \begin{align}
 I\left(s_{i,k}^{(i)};\mathbf{\hat{y}}_{i}^{(i)}\right) &\triangleq
f_{i,k,P}\left(\mathbf{\bar{L}}_{i},\mathbf{v}_{i}^{(i)}\right) \\
& = \mathrm{log}_{2}\left(1+ \big|v_{i,k}^{(i)}\big|^2 \mathbf{h}_{i,k}^{iH}\bar{\mathbf{L}}_{i}^{H} \mathbf{J}_{i,k}^{(i)\,-1}\bar{\mathbf{L}}_{i}\mathbf{h}_{i,k}^{i}\right). \nonumber
 \end{align}
 Here, the channel vector $\mathbf{h}_{i,k}^{i}$ is obtained by vertically stacking the vectors $\{\mathbf{h}_{i,k}^{i,r}\}_{r\in\mathcal{N}_{R,i}}$, and the matrices $\mathbf{\bar{L}}_{i}$ and $\mathbf{J}_{i,k}^{(i)}$ are defined as $\mathbf{\bar{L}}_{i} = \mathrm{diag}(\{\mathbf{L}_{i,r}^{(i)}\}_{r\in\mathcal{N}_{R,i}})$ and $\mathbf{J}_{i,k}^{(i)} = \sum_{l\in\mathcal{N}_{U,i}\backslash\{k\}}\lambda(\mathbf{\bar{L}}_{i}\mathbf{h}_{i,k}^{i}v_{i,l}^{(i)})+N_{0}\lambda(\mathbf{\bar{L}}_{i})+\mathbf{I}$, respectively.

Similarly, CP $i$ decodes the shared-subband messages $M_{i,k}^{(S)}$, $k\in\mathcal{N}_{U,i}$, by leveraging the quantized signals
$\mathbf{\hat{y}}_{i}^{(S)}=[\mathbf{\hat{y}}_{i,1}^{(S)H}\cdots$ $\mathbf{\hat{y}}_{i,N_{R,i}}^{(S)H}  \mathbf{\hat{y}}_{\bar{i},r_{\bar{i}},1}^{(S)H}\cdots \mathbf{\hat{y}}_{\bar{i},r_{\bar{i},S_{R,\bar{i}}}}^{(S)H}]^{H}$ of the shared subband that have been received from the proprietary RUs as well as from the other-tenant CP $\bar{i}$ on the backhaul link. Here, we have denoted the $S_{R,i}$ indices in $\mathcal{S}_{R,i}$ by $r_{i,1},r_{i,2},\ldots,r_{i,S_{R,i}}$.
We assume SUD detection, so that the achievable rate $R_{i,k}^{(S)}$ of message $M_{i,k}^{(S)}$ is given as $R_{i,k}^{(S)} = W_{S} I(s_{i,k}^{(S)};\mathbf{\hat{y}}_{i}^{(S)})$, where the mutual information is given as
\begin{align}
I\left(s_{i,k}^{(S)};\mathbf{\hat{y}}_{i}^{(S)}\right)& \triangleq f_{i,k,S}\left(\mathbf{\tilde{L}}_{i},\mathbf{v}_{i}^{(S)},\mathbf{v}_{\bar{i}}^{(S)}\right) \label{eq:mutual-shared-rate}\\
& = \mathrm{log}_{2}\left(1+ \big|v_{i,k}^{(S)}\big|^2 \mathbf{\tilde{h}}_{i,k}^{H}\tilde{\mathbf{L}}_{i}^{H}\mathbf{J}_{i,k}^{(S)\,-1}\tilde{\mathbf{L}}_{i}\mathbf{\tilde{h}}_{i,k}\right), \nonumber
\end{align}
with the channel vectors $\mathbf{\tilde{h}}_{i,k}$ and $\mathbf{\tilde{g}}_{i,k}$, that vertically stack the vectors $\{\mathbf{h}_{i,k}^{{i,r}}\}_{r\in\mathcal{N}_{R,i}}, \{\mathbf{h}_{i,k}^{\bar{i},r}\}_{r\in\mathcal{S}_{R,\bar{i}}}$ and $\{ \mathbf{h}_{i,k}^{\bar{i},r} \}_{r\in\mathcal{N}_R}, \{ \mathbf{h}_{i,k}^{i,r} \}_{r\in\mathcal{S}_{R,\bar{i}}}$, respectively, and the notations $\mathbf{\tilde{L}}_{i}=\mathrm{diag}( \{\mathbf{L}_{i,r}^{(S)}\}_{r\in\mathcal{N}_{R,i}}, \{\mathbf{L}_{\bar{i},r}^{(S)}\}_{r\in\mathcal{S}_{R,\bar{i}}} )$ and  $\mathbf{J}_{i,k}^{(S)}  =\sum_{l\in\mathcal{N}_{U,i}\backslash\{k\}}\lambda(\mathbf{\tilde{L}}_{i}\mathbf{\tilde{h}}_{i,l}v_{i,l}^{(S)})+\sum_{l\in\mathcal{N}_{U,\bar{i}}}\lambda(\mathbf{\tilde{L}}_{i}\mathbf{\tilde{g}}_{\bar{i},l}v_{\bar{i},l}^{(S)})
 + N_{0}\lambda(\mathbf{\tilde{L}}_{i})+\mathbf{I}$.

\subsection{Limiting Inter-Operator Privacy Loss \label{subsec:privacy}}

Exchange of the quantized signals on the backhaul links may cause a privacy loss among the participating operators. In fact, CP $\bar{i}$ can partially infer the shared-subband messages $\{M_{i,k}^{(S)}\}_{k\in\mathcal{N}_{U,i}}$ of the UEs of operator $i$ from the signals $\{\hat{\mathbf{y}}_{i,r}^{(S)}\}_{r\in\mathcal{S}_{R,i}}$ received on the backhaul link.

A way of limiting the privacy loss is to design the bandwidth allocation  $\mathbf{W}$, the power control $\mathbf{v}=\{\mathbf{v}_{i}^{(m)}\}_{i\in\mathcal{N}_{O}, m\in\{i,S\}}$ and the fronthaul quantization strategies $\mathbf{L}=\{\mathbf{L}_{i,r}^{(m)}\}_{i\in\mathcal{N}_{O}, r\in\mathcal{N}_{R,i},  m\in\{i,S\}}$ while imposing the information theoretic privacy constraint \cite{Park-et-al:TVT18}
\begin{equation}
\label{eq:privacy-constraint}
W_{S}\beta_{i,k,S}\left(\mathbf{\tilde{L}}_{\bar{i}},\mathbf{v}_{\bar{i}}^{(S)},\mathbf{v}_{i}^{(S)}\right) \leq \Gamma,
\end{equation}
where the function $\beta_{i,k,S}(\mathbf{\tilde{L}}_{\bar{i}},\mathbf{v}_{\bar{i}}^{(S)},\mathbf{v}_{i}^{(S)})$, that measures the amount information leaked to CP $\bar{i}$ per unit bandwidth, is defined as the mutual information
\begin{align}
& \beta_{i,k,S}\left(\mathbf{\tilde{L}}_{\bar{i}},\mathbf{v}_{\bar{i}}^{(S)},\mathbf{v}_{i}^{(S)}\right) \triangleq I\left(\mathbf{x}_{i,k}^{(S)};\mathbf{\hat{y}}_{\bar{i}}^{(S)}\right)\label{eq:mutual-privacy}\\
& =\mathrm{log}_{2}\mathrm{det}\left(\begin{array}{c}\sum_{l\in\mathcal{N}_{U,\bar{i}}}\lambda\left(\tilde{\mathbf{L}}{}_{\bar{i}}\tilde{\mathbf{h}}{}_{\bar{i},l}v_{\bar{i},l}^{(S)}\right)+N_{0}\lambda\left(\tilde{\mathbf{L}}{}_{\bar{i}}\right)\\+\sum_{l\in\mathcal{N}_{U,i}}\lambda\left(\tilde{\mathbf{L}}{}_{\bar{i}}\tilde{\mathbf{g}}{}_{i,l}v_{i,l}^{(S)}\right)+\mathbf{I}\end{array}\right) \nonumber \\
& -  \mathrm{log}_{2}\mathrm{det}\left(\begin{array}{c}\sum_{l\in\mathcal{N}_{U,\bar{i}}}\lambda\left(\tilde{\mathbf{L}}{}_{\bar{i}}\tilde{\mathbf{h}}{}_{\bar{i},l}v_{\bar{i},l}^{(S)}\right)+N_{0}\lambda\left(\tilde{\mathbf{L}}{}_{\bar{i}}\right)\\+\sum_{l\in\mathcal{N}_{U,i}\backslash\{k\}}\lambda\left(\tilde{\mathbf{L}}{}_{\bar{i}}\tilde{\mathbf{g}}{}_{i,l}v_{i,l}^{(S)}\right)+\mathbf{I}\end{array}\right) \nonumber,
\end{align}
and $\Gamma$ represents the threshold value for the amount of privacy loss.
A similar information-theoretic privacy constraint was considered in \cite{Park-et-al:TVT18} for the design of a multi-tenant C-RAN downlink system.

\section{Optimization}\label{sec:optimization}

In this section, we address the joint optimization of the rate variables $\mathbf{R}=\{R_{i,k}^{(m)}\}_{i\in\mathcal{N}_{O},  k\in{\mathcal{N}}_{U,i}, m\in\{i,S\}}$, the bandwidth allocation $\mathbf{W}$, the power control $\mathbf{v}$ and the fronthaul quantization strategies $\mathbf{L}$ with the goal of maximizing the sum-rate $R_{\mathrm{sum}} = \sum_{i\in\mathcal{N}_O, k\in\mathcal{N}_{U,i}, m\in\{i,S\}} R_{i,k}^{(m)}$ of the participating operators' UEs  under the constraints on the transmit power, fronthaul and backhaul capacity and the inter-operator privacy levels. The problem is formulated as
\begin{subequations}\label{eq:problem-original}
\begin{align}
&\!\!\!\!\underset{\mathbf{v},\mathbf{L},\mathbf{W},\mathbf{R}}{\mathrm{maximize}}\,\,  R_{\mathrm{sum}}\label{eq:problem-original-objective}\\
&\,\,\mathrm{s.t.}\,\, R_{i,k}^{(i)}\!\leq\! W_{P,i}f_{i,k,P}(\mathbf{\bar{L}}_{i},\mathbf{v}_{i}^{(i)}), \,i\in\mathcal{N}_{O}, k\in\mathcal{N}_{U,i}, \label{eq:problem-original-private-rate}\\
 & \,\,\,\,\,\,\,\,\,\,\,\,\, R_{i,k}^{(S)}\leq W_{S}f_{i,k,S}(\mathbf{\tilde{L}}_{i},\mathbf{v}_{i}^{(S)},\mathbf{v}_{\bar{i}}^{(S)}), \, i\in\mathcal{N}_{O},\,k\in\mathcal{N}_{U,i}, \label{eq:problem-original-shared-rate}\\
 & \,\,\,\,\,\,\,\,\,\,\,\,\, W_{P,i}g_{i,r}^{(i)}(\mathbf{L}_{i,r}^{(i)},\mathbf{v}_{i}^{(i)})+W_{S}g_{i,r}^{(S)}(\mathbf{L}_{i,r}^{(S)},\mathbf{v}_{i}^{(S)},\mathbf{v}_{\bar{i}}^{(S)}),\nonumber \\
 & \qquad \quad \leq C_{F,i,r},\,i\in\mathcal{N}_{O},\,r\in\mathcal{N}_{R,i},\label{eq:problem-original-fronthaul}\\
 & \,\,\,\,\,\,\,\,\,\,\,\,\, W_{S}\sum\nolimits_{r\in\mathcal{S}_{R,i}} g_{i,r}^{(S)}(\mathbf{L}_{i,r}^{(S)},\mathbf{v}_{i}^{(S)},\mathbf{v}_{\bar{i}}^{(S)})  \leq C_{B,i},\!i\!\in\!\mathcal{N}_{O}, \label{eq:problem-original-backhaul}\\
 & \,\,\,\,\,\,\,\,\, W_{S}\beta_{i,k,S}(\mathbf{\tilde{L}}_{\bar{i}},\mathbf{v}_{\bar{i}}^{(S)},\mathbf{v}_{i}^{(S)})\leq\Gamma, \, i\in\mathcal{N}_{O},\,k\in\mathcal{N}_{U,i}, \label{eq:problem-original-privacy}\\
 & \,\,\,\,\,\,\,\,\,\,\,\,\,\big|\mathbf{v}_{i,k}^{(m)}\big|^2 \leq P_{\max}, \,i\in\mathcal{N}_{O}, \,k\in\mathcal{N}_{U,i},\, m\in\{i,S\},
  \label{eq:problem-original-power}\\
 & \,\,\,\,\,\,\,\,\,\,\,\,\, \sum\nolimits_{i\in\mathcal{N}_{O}}W_{P,i}+W_{S}=W.\label{eq:problem-original-BW}
\end{align}
\end{subequations}

The problem (\ref{eq:problem-original}) is non-convex due to the constraints (\ref{eq:problem-original-private-rate})-(\ref{eq:problem-original-privacy}). To find an efficient solution, we adopt the matrix FP approach proposed in \cite{Shen-et-al:TN19}. Specifically, by using the results of \cite[Prop. 1, Cor. 1]{Shen-et-al:TN19} and the Fenchel conjugate function \cite{Borwein06}, we can show that the following conditions are stricter than (\ref{eq:problem-original-private-rate}), (\ref{eq:problem-original-backhaul}) and (\ref{eq:problem-original-privacy}):
\begin{align}
& R_{i,k}^{(i)} \leq 2c_{i,k,P} W_{P,i}^{1/2} - c_{i,k,P}^2 / \alpha_{i,k,P}, \nonumber\\
& \alpha_{i,k,P} \leq \log_2(1 + \kappa_{i,k,P}) - \frac{\kappa_{i,k,P}}{\ln 2} + \frac{1 + \kappa_{i,k,P}}{\ln 2} \times \nonumber \\
& \!\!\left( 2 v_{i,k}^{(i)H}\mathbf{h}_{i,k}^{iH} \bar{\mathbf{L}}_i^H\mathbf{z}_{i,k,P} - \mathbf{z}_{i,k,P}^H \! \left(\! \lambda(\bar{\mathbf{L}}_i \mathbf{h}_{i,k}^i v_{i,k}^{(i)}) +  \mathbf{J}_{i,k}^{(i)}\right)\! \mathbf{z}_{i,k,P} \! \right)\!, \nonumber \\
& \sum\nolimits_{r\in\mathcal{S}_{R,i}} \rho_{i,r}^{(S)} \leq C_{B,i}, \,\,\,\, \tilde{\rho}_{i,r}^{(S)} \leq 2 \tilde{c}_{i,r}^{(S)} \rho_{i,r}^{(S)1/2} \tilde{c}_{i,r}^{(S)2} W_S,  \nonumber \\
& \tilde{\rho}_{i,r}^{(S)} \geq \log_2\det (\mathbf{\Sigma}_{i,r}^{(S)}) + n_{R,i,r} / \ln 2 + 1/\ln 2 \times \nonumber \\
&
\mathrm{tr}\!\!\left(\! \mathbf{\Sigma}_{i,r}^{(S)\, -1} \!\!\left( \sum_{j\in\mathcal{N}_O, l\in\mathcal{N}_{U,j}} \!\!\!\!\! \lambda( \mathbf{L}_{i,r}^{(S)} \mathbf{h}_{j,l}^{i,r} v_{j,l}^{(S)} ) \! + \! N_o \lambda(\mathbf{L}_{i,r}^{(S)}) + \mathbf{I}  \! \! \right) \!\! \right) \!, \nonumber \\
& \beta_{i,k,S} \leq 2 \hat{c}_{i,k,S} \gamma^{1/2} - \hat{c}_{i,k,S}^2 W_S, \nonumber \\
& \beta_{i,k,S}\! \geq -\theta_{i,k,S} + \log_2\det(\tilde{\mathbf{\Sigma}}_i^{(S)}) \! - \! \frac{ n_{R,\bar{i}} + \sum\nolimits_{r\in \mathcal{S}_{R,i}} \! n_{R,i,r} }{\ln 2} \nonumber \\
& \!+\! \frac{1}{\ln 2} \mathrm{tr}\!\left( \! \tilde{\mathbf{\Sigma}}_i^{(S)-1} \!\! \left( \!\! \begin{array}{c} N_0\lambda(\tilde{\mathbf{L}}_{\bar{i}}) + \sum_{l\in\mathcal{N}_{U,\bar{i}}} \! \lambda(\tilde{\mathbf{L}}_{\bar{i}}\tilde{\mathbf{h}}_{\bar{i},l}v_{\bar{i},l}^{(S)})  \\  + \! \sum_{l\in\mathcal{N}_{U,i}} \lambda(\tilde{\mathbf{L}}_{\bar{i}} \tilde{\mathbf{g}}_{i,l}v_{i,l}^{(S)} ) + \mathbf{I}  \end{array}  \!\! \right) \! \right), \nonumber \\
& \theta_{i,k,S} \leq \log_2\det(\mathbf{I} + \mathbf{K}_{i,k}) - \mathrm{tr} (\mathbf{K}_{i,k}) / \ln 2 +  \nonumber \\
& \,\,\, \mathrm{tr}\!\left(\! (\mathbf{I} + \mathbf{K}_{i,k}) \! \left( 2\mathbf{A}_{\bar{i},k} \mathbf{Z}_{i,k} - \mathbf{Z}_{i,k}^H (\mathbf{A}_{\bar{i},k}^H \mathbf{A}_{\bar{i},k} + \mathbf{I}) \mathbf{Z}_{i,k} \right) \! \right) / \ln 2, \nonumber
\end{align}
for $i\in\mathcal{N}_O$, $k\in\mathcal{N}_{U,i}$ and $r\in\mathcal{N}_{R,i}$, where the matrix $\mathbf{A}_{\bar{i},k}$ is obtained by horizontally stacking the vectors $\{\tilde{\mathbf{L}}_{\bar{i}} \tilde{\mathbf{h}}_{\bar{i},l}v_{\bar{i},l}^{(S)}\}_{l\in\mathcal{N}_{U,\bar{i}}}$, $\{ \tilde{\mathbf{L}}_{\bar{i}} \tilde{\mathbf{g}}_{i,l} v_{i,l}^{(S)} \}_{l\in\mathcal{N}_{U,i}\setminus \{k\}}$ and the matrix $N_0^{1/2} \tilde{\mathbf{L}}_{\bar{i}}$. By setting the auxiliary variables $\kappa_{i,k,P}$, $\mathbf{z}_{i,k,P}$, $\tilde{c}_{i,r}^{(S)}$, $\mathbf{\Sigma}_{i,r}^{(S)}$, $\hat{c}_{i,k,S}$, $\tilde{\mathbf{\Sigma}}_i^{(S)}$, $\mathbf{K}_{i,k}$ and $\mathbf{Z}_{i,k}$ as
\begin{align}
& \kappa_{i,k,P} = v_{i,k}^{(i)H}\mathbf{h}_{i,k}^{iH} \bar{\mathbf{L}}_i^H \mathbf{J}_{i,k}^{(i)-1} \bar{\mathbf{L}}_i \mathbf{h}_{i,k}^i v_{i,k}^{(i)}, \nonumber \\
& \mathbf{z}_{i,k,P} = \left( \lambda(\bar{\mathbf{L}}_i\mathbf{h}_{i,k}^i v_{i,k}^{(i)}) + \mathbf{J}_{i,k}^{(i)} \right)^{-1} \bar{\mathbf{L}}_i \mathbf{h}_{i,k}^i v_{i,k}^{(i)}, \nonumber \\
& \tilde{c}_{i,r}^{(S)} = \rho_{i,r}^{(S)1/2} / W_S, \,\,\, \hat{c}_{i,k,S} = \Gamma^{1/2} / W_S, \nonumber \\
& \mathbf{\Sigma}_{i,r}^{(S)} = \sum_{j\in\mathcal{N}_O, l\in\mathcal{N}_{U,j}} \lambda(\mathbf{L}_{i,r}^{(S)} \mathbf{h}_{j,l}^{i,r} v_{j,l}^{(S)}) + N_0 \lambda(\mathbf{L}_{i,r}^{(S)}) + \mathbf{I},  \nonumber \\
%& \hat{c}_{i,k,S} = \Gamma^{1/2} / W_S, \nonumber \\
& \tilde{\mathbf{\Sigma}}_i^{(S)} = \sum_{l\in\mathcal{N}_{U,\bar{i}}} \lambda(\tilde{\mathbf{L}}_{\bar{i}} \tilde{\mathbf{h}}_{\bar{i},l}v_{\bar{i},l}^{(S)}) + \sum_{l\in\mathcal{N}_{U,i}} \lambda( \tilde{\mathbf{L}}_{\bar{i}} \tilde{\mathbf{g}}_{i,l} v_{i,l}^{(S)} ) \nonumber \\
& \,\,\,\,\,\,\,\,\,\,\,\,\,\,\,\,\,\,\,\,\, + N_0\lambda(\tilde{\mathbf{L}}_{\bar{i}}) + \mathbf{I}, \nonumber \\
& \mathbf{K}_{i,k} = \mathbf{A}_{\bar{i},k}\mathbf{A}_{\bar{i},k}^H, \,\,\, \mathbf{Z}_{i,k} = \left( \mathbf{A}_{\bar{i},k}^H\mathbf{A}_{\bar{i},k} + \mathbf{I} \right)^{-1} \mathbf{A}_{\bar{i},k}^H, \nonumber %\\
%& \mathbf{Z}_{i,k} = \left( \mathbf{A}_{\bar{i},k}^H\mathbf{A}_{\bar{i},k} + \mathbf{I} \right)^{-1} \mathbf{A}_{\bar{i},k}^H, \nonumber
\end{align}
it can be seen that the inequalities above are in fact equivalent to (\ref{eq:problem-original-private-rate}), (\ref{eq:problem-original-backhaul}), and (\ref{eq:problem-original-privacy}) \cite{Shen-et-al:TN19,Borwein06}.

The remaining non-convex constraints (\ref{eq:problem-original-shared-rate}) and (\ref{eq:problem-original-fronthaul}) can be similarly converted to stricter conditions, that become equivalent to (\ref{eq:problem-original-shared-rate})-(\ref{eq:problem-original-fronthaul}) when auxiliary variables are properly selected as given closed-form expressions.
The proposed algorithm tackles the alternative problem obtained by replacing the non-convex constraints (\ref{eq:problem-original-private-rate})-(\ref{eq:problem-original-privacy}) with the discussed stricter, but convex, constraints. The obtained problem with respect to any one of the variables $\mathbf{W}$, $\mathbf{v}$ and $\mathbf{L}$, along with $\mathbf{R}$, while fixing the other variables and the auxiliary variables, is convex. Therefore, it can be addressed via a standard convex solver. For fixed variables $\mathbf{W}$, $\mathbf{v}$ and $\mathbf{L}$, the proposed scheme then updates the auxiliary variables so that the stricter conditions match the original constraints as discussed above.
The resulting iterative algorithm, which alternately updates the variables $\{\mathbf{W}, \mathbf{v}, \mathbf{L}\}$ and the auxiliary variables, achieves monotonically non-decreasing objective values.

\section{Numerical Results}\label{sec:results}

This section shows numerical results to verify the advantages of the proposed multi-tenant C-RAN uplink system. We assume that the UEs and RUs are uniformly distributed within a circular region of radius $100$ m.
For the elements of the channel vectors, we assume independent and identically distributed (i.i.d.) Rayleigh fading with the path-loss  $1/(1+(d/d_{\mathrm{ref}})^{\alpha})$. Here, $d$ denotes the distance, and we set $d_{\mathrm{ref}}=50$ m and $\alpha=3$ in the simulation.

\begin{figure}
\centering
\centerline{\includegraphics[width=8.4cm, height=6.5cm]{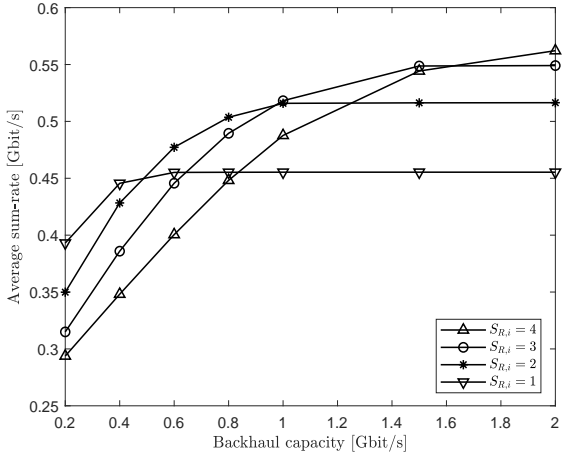}}
\caption{ {\scriptsize Average sum-rate versus backhaul capacity ($N_{R,i}= N_{U,i}= 4$, $n_{R,i,r}=1$, $C_{F,i,r}= 500 \,\mathrm{Mbps}$, $W=100 \,\mathrm{MHz}$, $\Gamma = 600$ Mbps and 0 dB SNR).} }
\label{fig:Backhaul}
\end{figure}

In Fig. \ref{fig:Backhaul}, we observe the impact of the number $S_{R,i}$ of streams delivered on the backhaul link for inter-operator cooperation by  plotting the average sum-rate of the proposed optimized spectrum pooling scheme with $S_{R,i}\in\{1,2,3,4\}$ versus the backhaul capacity $C_{B,i}$ for the uplink of a multi-tenant C-RAN system with $N_{R,i}= 4$, $N_{U,i}= 4$, $n_{R,i,r}=1$, $C_{F,i,r}= 500 \,\mathrm{Mbps}$, $W=100 \,\mathrm{MHz}$, $\Gamma = 600$ Mbps and 0 dB SNR.
The figure shows that, for small backhaul capacity, it is desirable to reduce the burden on the backhaul links by decreasing $S_{R,i}$. In contrast, for sufficiently large backhaul capacity and privacy threshold $\Gamma$, it is preferable to exchange as much information as possible on the backhaul links in order to maximize the impact of inter-operator cooperation.

\begin{figure}
\centering
\centerline{\includegraphics[width=8.4cm, height=6.5cm]{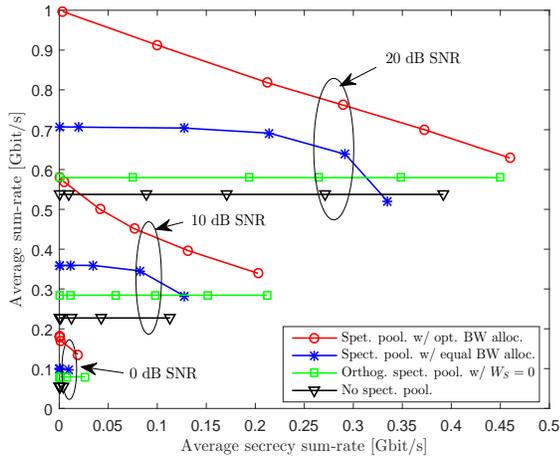}}
\caption{ {\scriptsize Average sum-rate versus average secrecy sum-rate ($N_{R,i} = N_{U,i}= 2$, $S_{R,i}= 2$, $n_{R,i,r}= 1$, $C_{B,i}= 1 \,\mathrm{Gbps}$, $C_{F,i,r}= 500 \,\mathrm{Mbps}$, $W=100 \,\mathrm{MHz}$, and 0, 10 and 20 dB SNRs).} }
\label{fig:SRvsSecrecy}
\end{figure}

In Fig. \ref{fig:SRvsSecrecy}, we plot the average sum-rate versus the average secrecy sum-rate for the uplink of a multi-tenant C-RAN system with $N_{R,i}= 2$, $N_{U,i}= 2$, $S_{R,i}= 2$, $n_{R,i,r}=1$, $C_{B,i}= 1 \,\mathrm{Gbps}$, $C_{F,i,r}= 500 \,\mathrm{Mbps}$, $W=100 \,\mathrm{MHz}$, and 0, 10 and 20 dB SNRs. We define the secrecy sum-rate as $R_{\mathrm{sum, sec}} = \sum_{i\in\mathcal{N}_{O}, k\in\mathcal{N}_{U,i}} R_{i,k, \mathrm{sec}}$, where $R_{i,k, \mathrm{sec}} = \max\{R_{i,k} - \Gamma, 0\}$
 measures the information rate at which UE $(i,k)$ can communicate with the CP $i$ without being eavesdropped by the other tenant CP $\bar{i}$.
We compare the performance of the proposed optimized spectrum pooling scheme with the following baseline schemes: $(i)$ No spectrum pooling with $W_{P,1} = W_{P,2} = W/2$ and $W_S=0$; $(ii)$ Spectrum pooling with equal bandwidth allocation: $W_{P,1} = W_{P,2} = W_S = W/3$; $(iii)$ Orthogonal spectrum pooling with optimized $W_{P,1}$ and $W_{P,2}$ for fixed $W_S=0$.
The figure shows that the proposed spectrum pooling scheme with optimized bandwidth allocation achieves a better trade-off between rate and privacy than the other schemes, and that the percentage gain of the  proposed scheme in terms of the sum-rate decreases with the secrecy rate. This suggests that, in order to guarantee a high level of privacy among the operators, the usage of the shared subband and the backhaul links should be minimized.
We also note that the performance of the spectrum pooling scheme with a fixed bandwidth $W_S = W/3$ of the shared subband can be worse than that of no spectrum pooling or of orthogonal spectrum pooling, particularly in the regime of large secrecy sum-rates.

\section{Conclusion}\label{sec:conclusion}

We have studied the advantages of spectrum pooling and inter-tenant cooperative reception for the uplink of a multi-tenant C-RAN system. For inter-tenant cooperation on the backhaul links, we proposed a cooperative reception strategy whereby CPs exchange subsets of the bit streams received on the fronthaul links without any processing. We tackled the problem of jointly optimizing the bandwidth allocation, the UE power control, and the fronthaul quantization strategies with the goal of maximizing the sum-rate of all the participating UEs, while taking into account the constraints on the inter-operator privacy loss.
We validated the performance gains of the proposed optimized spectrum pooling scheme via numerical results.
As open research issue, we mention the design of in-network processing strategy \cite[Sec. IV]{Park-et-al:TVT16} for inter-operator cooperation on the backhaul links.

\bibliographystyle{IEEEbib}
\bibliography{reference}
%\bibliography{multiTenantCRAN_UL_conf}

\end{document}